\documentclass[twocolumn,prl, amsmath,amssymb,aps]{revtex4}
\usepackage{graphicx}
\usepackage{titlesec}
\usepackage{color}
\begin{document}

\title{Electrical Detection of Direct and Alternating Spin Current Injected from a Ferromagnetic Insulator into a Ferromagnetic Metal}
\author{P. Hyde, Lihui Bai, D.M.J. Kumar, B.W. Southern, and C.-M. Hu}\email{hu@physics.umanitoba.ca}
\affiliation{Department of Physics and Astronomy, University of Manitoba, Winnipeg, Canada R3T 2N2}
\author{S. Y. Huang, B. F. Miao, and C. L. Chien}
\affiliation{Department of Physics and Astronomy, Johns Hopkins University, Baltimore, MD 21218, USA}
\date{\today}
\begin{abstract}
We report room temperature electrical detection of spin injection from a ferromagnetic insulator (YIG) into a ferromagnetic metal (Permalloy, Py). Non-equilibrium spins with both static and precessional spin polarizations are dynamically generated by the ferromagnetic resonance of YIG magnetization, and electrically detected by Py as dc and ac spin currents, respectively. The dc spin current is electrically detected via the inverse spin Hall effect of Py, while the ac spin current is converted to a dc voltage via the spin rectification effect of Py which is resonantly enhanced by dynamic exchange interaction between the ac spin current and the Py magnetization. Our results reveal a new path for developing insulator spintronics, which is distinct from the prevalent but controversial approach of using Pt as the spin current detector.

\end{abstract}

\maketitle

Developing new methods for generating and detecting spin currents has been the central task of spintronics. In the pioneering work of Johnson and Silsbee \cite{Johnson1985PRL}, the generation and detection of spin-polarized currents were both achieved through the use of ferromagnetic metals (FM). Recent breakthroughs reveal ferromagnetic insulators (FI) to be promising spin current sources, in which spin currents can be generated without the presence of any charge current \cite{Kajiwara2010Nat,Uchida2010APL}. In the ground-breaking experiment performed 3 years ago by Kajiwara \textit{et al.} \cite{Kajiwara2010Nat}, electrical detection of the spin current generated by yttrium iron garnet (Y$_{3}$Fe$_{5}$O$_{12}$, YIG) was achieved by utilizing the heavy normal metal platinum (Pt), in which spin current was detected via the inverse spin Hall effect (ISHE). Since then, nearly the entire insulator-spintronics community has followed suit and used Pt as the standard spin detector. But so far, consensus has not yet been achieved on a few critical spin-dependent material issues of Pt \cite{ Huang2011PRL, Huang2012PRL, Qu2013PRL, Kikkawa2013PRL}. Given the fact that ferromagnetic metals are broadly used as spin detectors in both semiconductor \cite{Hu2001PRB, Luo2007NP} and metallic spintronics devices \cite{Johnson1985PRL,Jedema2001Nat, Jedema2002Nat}, it is noteworthy that the appealing topic of how a FM material may detect the spin current generated by a FI has barely been investigated. Elucidating this issue is of broad interest for making insulator-spintronics device compatible with both semiconductor and metallic spintronics devices.

In this letter, we report room temperature detection of spin current generated in YIG by feromagnetic resonance (FMR). Distinct from the popular approach of using Pt as the spin detector, we use the ferromagnetic metal Permalloy (Py) instead, and demonstrate that Py not only detects the dc spin current from YIG, but most strikingly, it also detects the recently predicted ac spin current \cite{Jiao2013PRL} by directly converting it into a dc voltage, which makes Py a superior spin detector compared to Pt. Two very recent experiments make this work possible: (i) the discovery of the ISHE in Py \cite{Miao2013PRL}, and (ii) the establishment of a universal method for clearly separating spin rectification from spin pumping \cite{Bai2013PRL}.

\begin{figure}[h!]
    \includegraphics[width = 8 cm]{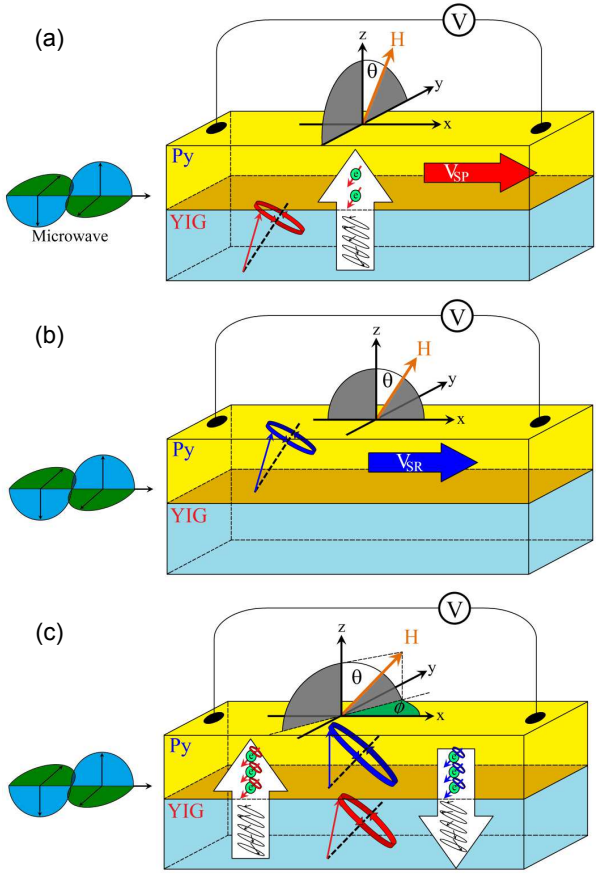}
    \caption{(Colour online) (a) At $\phi$=90$^{\circ}$, the dc spin current pumped by YIG FMR can be detected in Py via the ISHE induced dc voltage $V_{\textmd{SP}}$. (b) At $\phi$=0$^{\circ}$, the Py FMR can be detected by $V_{\textmd{SR}}$ via the spin rectification effect. (c) At the equal-resonance condition, the ac spin current pumped from the YIG enhances the FMR of Py, which can be detected by the increased $V_{\textmd{SR}}$. Correspondingly, enhanced YIG FMR can be detected via the increased $V_{\textmd{SP}}$.}
    \label{fig_sketch}
\end{figure}

We begin by highlighting the basic ideas. As shown in Fig. \ref{fig_sketch}, let us consider a Py/YIG bilayer under microwave irradiation in an external magnetic field {\bf H}. Choosing the $x$ axis as the longitudinal direction for measuring the dc voltages, and the $z$ axis as perpendicular to the interface, the direction of {\bf H} is described by the polar (with respect to the $z$ axis) and azimuth (with respect to the $x$ axis) angles of $\theta$ and $\phi$, respectively, as shown in Fig. \ref{fig_sketch}(c). At the FMR frequency $\omega_{\textmd{YIG}}$ of YIG, the magnetization of YIG precesses about its saturation magnetization $\bf{M}$, which pumps non-equilibrium spins diffusing across the Py/YIG interface. Hence, a dc spin current $\bf j_s$ carries static non-equilibrium spin angular momentum which is antiparallel to $\bf M$, while an ac spin current $\bf j_s(\omega_{YIG})$ carries dynamic non-equilibrium spin angular momentum which is precessing about $\bf M$ \cite{Jiao2013PRL}. Both spin currents flow along the $z$ direction, as shown in Fig. \ref{fig_sketch}(c).

Based on the recently discovered ISHE in Py \cite{Miao2013PRL}, the idea of using Py to detect $\bf j_s$ is straightforward as shown in Fig. \ref{fig_sketch}(a). It can be detected by the dc voltage $V_{\textmd{SP}}$ in Py produced through spin pumping and the ISHE, \textit{i.e.}, $V_{\textmd{SP}}$ is proportional to $j_s$. In contrast, detecting the high-frequency ac spin current $\bf j_s(\omega_{YIG})$ is nontrivial and is currently of great interest. Two groups have very recently developed very smart methods to solve this problem \cite{Wei2013arxiv, Hahn2013arxiv}. Both use a microwave detector for measuring the ac current in Pt induced by $\bf j_s(\omega_{YIG})$. Different from the two methods \cite{Wei2013arxiv, Hahn2013arxiv}, our idea is inspired by the pioneering work of the forgotten masters Silsbee \textit{et al.}, who performed 35 years ago the first spin pumping experiment via the enhanced spin resonance \cite{Silsbee1979PRB}. And we utilize the spin rectification effect in Py which we have systematically studied \cite{Gui2007PRL, Mecking2007PRB, Wirthmann2010PRL, Harder2011PRB}. At the Py FMR frequency $\omega_{\textmd{Py}}$, the precessing magnetization leads to the spin rectification which induces a dc voltage $V_{\textmd{SR}}$ proportional to the precession angle, as shown in Fig. \ref{fig_sketch}(b). At the equal-resonance condition where $\omega_{\textmd{YIG}}$ = $\omega_{\textmd{Py}}$ [shown in Fig. \ref{fig_sketch}(c)] the ac spin current precessing at $\omega_{\textmd{YIG}}$ may enhance the FMR of Py via dynamic exchange interaction, in a process similar to the enhanced electron spin resonances discovered by Silsbee \textit{et al.} \cite{Silsbee1979PRB}. Thus, measuring the enhanced $V_{\textmd{SR}}$ in Py may permit direct electrical detection of the ac spin current without the use of any microwave detectors.

Such a method needs two prerequisites: (i) a clear procedure for distinguishing $V_{\textmd{SP}}$ from $V_{\textmd{SR}}$, and (ii) a practical way for setting the equal-resonance condition where $\omega_{\textmd{YIG}}$ = $\omega_{\textmd{Py}}$ at the same magnetic field $H$, or equivalently, setting the FMR resonance field $H_{\textmd{YIG}}$ = $H_{\textmd{Py}}$ at the same microwave frequency $\omega$. The required procedure has recently been established \cite{Bai2013PRL} so that we may use the following angular condition and symmetries to clearly separate and identify the dc voltages induced by pure spin pumping ($V_{\textmd{SP}}$) and pure spin rectification ($V_{\textmd{SR}}$):
\begin{align}
\nonumber
At~\phi=90^{\circ}, V_{\textmd{SP}}(\theta, H)& = -V_{\textmd{SP}}(\theta, -H) = -V_{\textmd{SP}}(-\theta, H);\\
At~\phi=0^{\circ}~~,V_{\textmd{SR}}(\theta, H)& =~~ V_{\textmd{SR}}(\theta, -H) = -V_{\textmd{SR}}(-\theta, H).
\label{eq_VH}
\end{align}
The equal-resonance condition, as we demonstrate below, can be set by adjusting the $\textbf{H}$ field direction, making use of the different magnetic anisotropies of Py and YIG.

Samples were prepared by magnetron sputtering and patterned using a photo-lithography and liftoff technique. A 10-nm thick Py thin film was deposited on a YIG substrate (10 mm $\times$ 4 mm in area) and patterned into Hall bar structure with lateral dimensions of 5 mm $\times$ 0.2 mm. A 100-mW microwave was applied to excite FMR in the bilayer through a rectangular waveguide. By sweeping the {\bf H} field at a fixed microwave frequency, dc voltages induced by FMR were detected along the $x$ axis of the Hall bar using a lock-in amplification. Here, the microwave power was modulated at a frequency of 8.33 kHz.

\begin{figure}[t]
    \includegraphics[width = 8.2 cm]{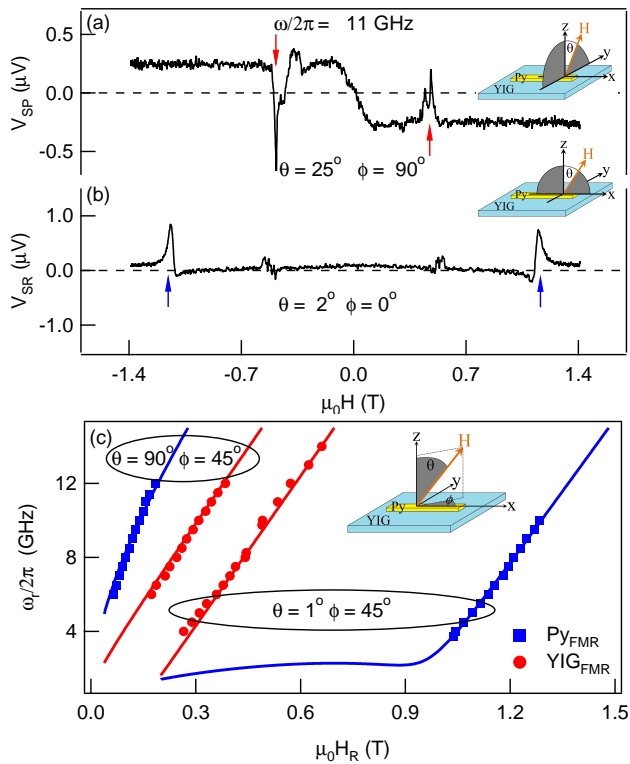}
    \caption{(Colour online) (a) The YIG and (b) the Py FMR electrically detected via $V_{\textmd{SP}}$ at $\phi=90^{\circ}$ and $V_{\textmd{SR}}$ at $\phi=0^{\circ}$, respectively. (c) $\omega_{r}-H_{R}$ dispersions of the Py and YIG FMRs measured at in-plane ($\theta$ = 90$^{\circ}$) and out-of-plane ($\theta \approx$ 0$^{\circ}$) field configurations. Curves are calculated theoretically.}
    \label{fig2_FMRs}
\end{figure}

Figure \ref{fig2_FMRs} shows typical voltage signals measured at $\omega/2\pi$ = 11 GHz. While sweeping the {\bf H} field applied at $\phi$ = 90$^{\circ}$, we observe a background signal of $\pm$0.3 $\mu$V and sharp resonances at $\mu_{0}H_{R}$ = $\pm$0.484 T with a line width of 10.0 mT as shown in Fig. \ref{fig2_FMRs}(a). At the lower (inner) field side of the sharp resonance, there is a weaker resonance together with a series of resonances too weak to be accurately distinguished. Both the background and resonance signals have an odd symmetry with respect to the {\bf H} field direction, \textit{i.e}, $V(H) = -V(-H)$. The data plotted in Fig. \ref{fig2_FMRs}(a) was taken at $\theta$ = 25$^{\circ}$, but data with an odd symmetry was measured at other angles of $\theta$ (not shown), provided $\phi$ = 90$^{\circ}$. In contrast, by setting $\phi$ = 0$^{\circ}$, both the background and the two sharp resonances nearly disappear, as shown in Fig. \ref{fig2_FMRs}(b). Instead, broader resonances at $\pm$1.137 T with a line width of 17.5 mT are observed, which have an asymmetric line shape but even field symmetry of $V(H) = V(-H)$. Again, as long as $\phi$ = 0$^{\circ}$, the broad resonances with even field symmetry are observed at arbitrary angle of $\theta$, but note that they do not appear in the spectrum measured at $\phi$ = 90$^{\circ}$.

Similar background voltage $V_{bg}$ has been found in other bilayer devices such as Pt/YIG under microwave excitation \cite{PC1}. In general, for devices with a thin metallic layer deposited on a thick substrate, microwave heating is known to cause a temperature gradient perpendicular to the interface \cite{Zhang2012PRL}. Hence, a simple interpretation is that such a vertical temperature gradient may drive a dc spin current via the spin Seebeck effect \cite{Kikkawa2013PRL}, which may be detected via the ISHE as $V_{bg}$. Indeed, we find that $V_{bg} \propto sin(\phi)$ as expected from the spin Seebeck effect. However, we note that such an angular dependence can not irrefutably rule out the possibility that $V_{bg}$ is caused by the anomalous Nernst effect \cite{Huang2012PRL} which leads to the same relation of $V_{bg} \propto sin(\phi)$. Hence, we leave the intriguing origin of $V_{bg}$ to a future study, and focus in this paper on the detection of spin currents via FMR, which can be conclusively verified.

When $\phi\ne$ $n\times\pi$/2 where $n$ is an integer, we find that both the sharp and broad resonances appear in the same voltage trace. Although their relative strength depends on $\phi$, as we have discussed, neither of their resonance fields is sensitive to this angle; both depend on the polar angle, $\theta$. Setting $\phi$ = 45$^{\circ}$, the dispersions for both resonances were measured at $\theta$ = 1$^{\circ}$ and 90$^{\circ}$, corresponding to perpendicular and in-plane $\bf H$ field directions, respectively. They are plotted in Fig. \ref{fig2_FMRs}(c) for comparison. To identify these resonances, we have calculated the FMR conditions for the Py/YIG bilayer by linearizing the Landau-Lifshitz-Gilbert equations about the equilibrium determined by the $\bf H$ field strength and direction. Because of the macroscopic lateral size of the device, we make the simplest approximation to model the magnetic anisotropy by using a perpendicular demagnetization field $\mu_{0}M_d$ as the fitting parameter. From the best fits we find $\mu_{0}M_d$ = 0.147 and 0.910 T for YIG and Py, respectively. The gyromagnetic factor is found to be $\gamma$ = 27.0 and 26.2 GHz/T for YIG and Py, respectively. Note that the thin Py film has a much larger perpendicular anisotropy than YIG, as expected.

The calculated dispersions are plotted in Fig. \ref{fig2_FMRs}(c) as solid curves. The good agreement allows us to identify the sharp and broad resonances in Fig. \ref{fig2_FMRs}(a) and (b) as the FMR of YIG and Py, respectively. Their different line widths are consistent with the fact that the damping constant of YIG is much smaller than that of Py. To keep the focus, our simple model includes neither the exchange coupling nor the high-order anisotropy of YIG, hence it does not explicitly explain the origin of the weak resonance in Fig. \ref{fig2_FMRs}(a), which could be the spin wave observed previously \cite{Kajiwara2010Nat}. Following Eq. \ref{eq_VH}, at $\phi$ = 90$^{\circ}$, the measured field symmetry of $V(H_{\textmd{YIG}}) \simeq -V(-H_{\textmd{YIG}})$ as shown in Fig. \ref{fig2_FMRs}(a) allows us to identify the dc voltage of the YIG FMR as $V_{\textmd{SP}}$ \cite{Bai2013PRL}. Hence, the dc spin current $\bf j_s$ injected from the YIG into the Py is electrically detected.

Similarly, at $\phi$ = 0$^{\circ}$, the measured field symmetry of $V(H_{\textmd{Py}}) \simeq V(-H_{\textmd{Py}})$ as shown in Fig. \ref{fig2_FMRs}(b) confirms that the Py FMR is electrically detected via pure spin rectification \cite{Gui2007PRL, Bai2013PRL}, which we now use to detect the ac spin current $\bf j_s(\omega_{YIG})$.

As shown in Fig. \ref{fig2_FMRs}(c), at the same microwave frequency, the Py FMR measured in the in-plane configuration with $\theta$ = 90$^{\circ}$ appears on the low field side of the YIG FMR. Due to the larger perpendicular anisotropy of Py, in the perpendicular configuration with $\theta$ = 1$^{\circ}$, the Py FMR moves to the high field side. Hence, the equal-resonance condition of Py and YIG can be set by tuning the polar angle $\theta$. With the obtained parameters we have calculated and found that it occurs at $\theta$ = 12$^{\circ}$.

\begin{figure}[t]
    \includegraphics[width = 8.2 cm]{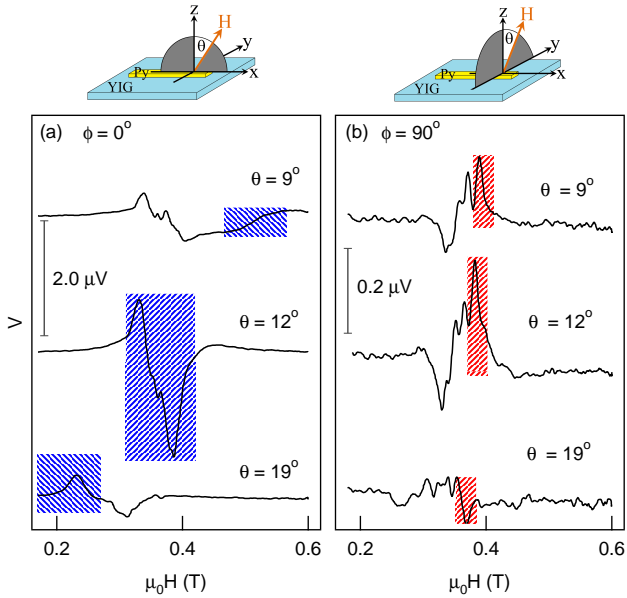}
    \caption{(Colour online) At $\theta=12^{\circ}$, both amplitudes of (a) the Py FMR measured by $V_{\textmd{SR}}$ ($\phi=0^{\circ}$) and (b) the YIG FMR measured by $V_{\textmd{SP}}$ ($\phi=90^{\circ}$) are greatly enhanced. In both (a) and (b) $\omega/2\pi$ = 7GHz.}
    \label{fig_SPvsSR}
\end{figure}

We thus proceeded to study the ac spin current enhanced FMR signal near $\theta$ = 12$^{\circ}$. Following Eq. \ref{eq_VH} by setting $\phi$ = 0$^{\circ}$, we can trace the electrically detected FMR of Py when $\theta$ is tuned through 12$^{\circ}$, as shown in Fig. \ref{fig_SPvsSR}(a). The shaded areas are the approximate calculated FMR fields of Py. When $\theta$ is tuned from 9$^{\circ}$ to 12$^{\circ}$, the peak-to-peak amplitude of the FMR signal is seen to increase by more than a factor of 4 (from below 0.5 $\mu$V to above 2 $\mu$V). When $\theta$ is further tuned from 12$^{\circ}$ to 19$^{\circ}$, the FMR signal amplitude drops back below 0.5 $\mu$V. Note that the detailed line shape of the FMR signal depends sensitively on the external field direction \cite{Harder2011PRB}, but at $\phi$ = 0$^{\circ}$ the amplitude of the FMR signal, electrically detected via spin rectification, provides a good measure of the cone angle of the magnetization precession \cite{Gui2007PRL}.

In order to rule out the possibility that the dramatically enhanced Py FMR signal is just due to the static interlayer exchange coupling \cite{Heinrich1990PRL}, we monitor the $\theta$ dependence of the YIG FMR signal detected by spin pumping at $\phi$ = 90$^{\circ}$. For the static coupling of Py and YIG magnetizations, one would only observe an anti-crossing of their FMRs, with the enhancement of one mode accompanied by the suppression of the other \cite{Heinrich1990PRL}. In contrast, as shown in Fig. \ref{fig_SPvsSR}(b), the FMR signal of YIG is also found to be enhanced dramatically when $\theta$ is tuned through 12$^{\circ}$.

\begin{figure}[t]
    \includegraphics[width = 8 cm]{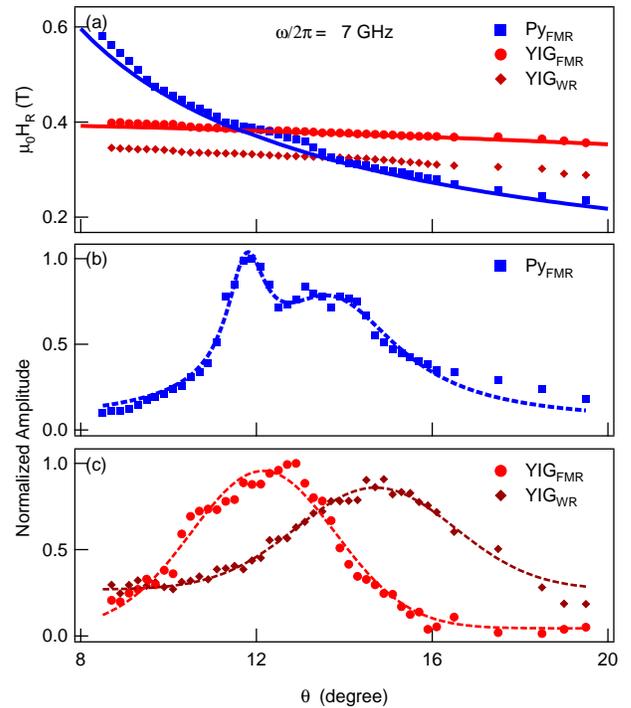}
    \caption{(Colour online) (a) The polar angular dependence of the resonance fields measured at $\omega/2\pi$ = 7GHz, showing the Py FMR crosses the YIG resonances at $\theta$ = 12$^{\circ}$ and 14$^{\circ}$. The normalized amplitudes of (b) the Py FMR and (c) the YIG resonances showing the simultaneous enhancement at equal-resonance conditions. Solid curves in (a) are calculated theoretically, dashed curves in (b) and (c) are guide to eyes.}
    \label{fig_coupling}
\end{figure}

Such a simultaneous enhancement of both FMR signals is more clearly seen from the systematic data measured at $\omega/2\pi $= 7 GHz. As shown in Fig. \ref{fig_coupling}(a), going from the perpendicular down to the in-plane configuration by increasing $\theta$, the FMR field of Py deceases much faster than that of YIG due to their different perpendicular anisotropies. It crosses first at $\theta$ = 12$^{\circ}$ with the YIG FMR (as calculated), then it crosses at about 14$^{\circ}$ with the weak resonance mode YIG$_{\textmd{WR}}$. Fig. \ref{fig_coupling}(b) shows the amplitude of Py FMR signal measured at $\phi$ = 0$^{\circ}$ via spin rectification, which is normalized by the maximum amplitude of 2.67 $\mu$V at $\theta$ = 12$^{\circ}$. For comparison, the amplitude of the YIG FMR measured at $\phi$ = 90$^{\circ}$ via spin pumping is plotted in Fig. \ref{fig_coupling}(c), which is normalized by the maximum amplitude of 0.35 $\mu$V, also at $\theta$ = 12$^{\circ}$. Clearly, at the equal-resonance condition, the amplitudes of both the Py and YIG FMR voltages increase dramatically and simultaneously.

It is intriguing to compare the simultaneously enhanced FMRs electrically detected in Py/YIG bilayer with the simultaneously narrowing of the FMRs measured by absorption spectroscopy on Fe/Au/Fe layers \cite{Heinrich2003PRL}. The absorption experiment performed by Heinrich \textit{et al.} is enlightening  since it reveals the exact cancellation of the spin currents flowing in opposite directions at equal-resonance condition, which reduces the damping of spin pumping. In our experiment, the dc voltage detected via the spin rectification effect measures the cone angle of Py FMR. At the equal-resonance condition, the ac spin current pumped by YIG FMR injects into Py, which reduces the damping and therefore enhances the cone angle of the Py FMR. In the phenomenological theory developed by Silsbee \textit{et al.} \cite{Silsbee1979PRB}, such an enhancement of spin resonance is caused by the dynamic exchange interaction between the ac spin current and the spin angular momentum. Either of these two pictures allow us to conclude that, by using the spin rectification of Py, the ac spin current of YIG can be electrically detected at the equal-resonance condition, as demonstrated in our experiment.

In summary, we have demonstrated new methods for the electrical detection of dc and ac spin currents in YIG. Both are achieved by using Py as the spin detector. Since the magnetization in Py is very easy to control by either tuning an external magnetic field or by tailoring its shape anisotropy, we expect that our straightforward methods permit the advancement of insulator spintronics in a distinct new path, setting it free from relying on Pt as the spin detector, in which the pivotal spin Hall effect is still controversial and is very difficult to tune.

Work in Manitoba has been funded by NSERC, CFI, URGP, and UMGF grants (C.-M.H. and B.W.S.). Work at JHU has been funded by NSF (DMR-1262253). D.M.J.K was supported by  Mitacs Globalink Program. S.Y.H was partially supported by STARnet sponsored by MARCO and DARPA.
 
\end{document}